\documentclass[article,twocolumn,
superscriptaddress,
amsmath, amssymb,
aps,
journal=prl
]{revtex4-2}

\usepackage{natbib}

\usepackage{hyperref}

\hypersetup{colorlinks=true,linkcolor=blue, citecolor=blue, urlcolor=blue, filecolor=blue}

\usepackage{graphicx}% Include figure files
\usepackage{dcolumn}% Align table columns on decimal point
\usepackage{bm}% bold math
\usepackage{xcolor}
\usepackage{float}
\usepackage{newtxtext,newtxmath}

\begin{document}

\title{Chiral Phonons Protected by Symmetry in a Strongly Anharmonic Helical Material}

\author{Supriyo Ghosh}
\email{supriyo@uchicago.edu}
\affiliation{Department of Chemistry, University of Chicago, Chicago, Illinois 60637, USA}

\author{Dipti Jasrasaria}
\email{djasrasaria@uchicago.edu}
\affiliation{Department of Chemistry, University of Chicago, Chicago, Illinois 60637, USA}
\affiliation{The James Franck Institute, University of Chicago, Chicago, Illinois 60637, USA}

\date{\today}

\begin{abstract}
How chiral phonons survive anharmonic phonon-phonon interactions, which underpin lattice dynamics at finite temperature, remains poorly understood. Using an \textit{ab initio} many-body approach, we study crystalline tellurium (Te), a strongly anharmonic helical crystal. We show that rotational symmetries in Te impose pseudoangular momentum selection rules on phonon-phonon interactions, restricting mode mixing and scattering at rotationally invariant symmetry points in the Brillouin zone. These symmetries protect the circular polarization of vibrational modes at finite temperature but not their lifetimes such that the dynamic chirality of handed vibrational modes persists to 600\,K, even as lifetimes drop sharply. We find that the temperature-dependence of angular momentum generated by a thermal gradient is governed by lifetimes rather than chirality, decreasing by two orders of magnitude between 10 and 600\,K. 
\end{abstract}

\maketitle

Chiral lattice vibrations, which exhibit well-defined handedness, have garnered significant interest over the past decade due to their ability to generate and exchange angular momentum~\cite{Juraschek2025, wang_chiral_2024}. These quasiparticles have been studied theoretically~\cite{zhang_angular_2014, zhang_chiral_2015, zhang_chiral_2022, hamada_phonon_2018, pols_chiral_2025, qin_chiral_2025, huang_2026}, verified experimentally~\cite{zhu_observation_2018, ishito_truly_2023, ueda_chiral_2023, Li_te_prl_2026}, and tied to a variety of phenomena, including magnet-free spin transport~\cite{qin_chiral_2025, kim_chiral-phonon-activated_2023,fransson_chiral_2023}, orbital angular momentum currents~\cite{Nabei2026, zhou_2026}, macroscopic magnetization~\cite{davies_phononic_2024, ferromagnetism, Tauchert2022}, and selective coupling to circularly polarized light~\cite{ishito_truly_2023, ishito_chiral_2023, lin_optically_2025}, highlighting the key role of chiral lattice vibrations in thermal management and quantum devices. While many experimental investigations of chiral lattice dynamics have been conducted in materials with soft lattices (\text{e.g.,} perovskites) and at elevated temperatures, theoretical characterizations of chiral lattice dynamics have been limited to harmonic phonon descriptions. As a result, the effects of temperature and anharmonic phonon-phonon interactions on chiral lattice dynamics are poorly understood.

\begin{figure*}[t]
    \includegraphics[width=\textwidth]{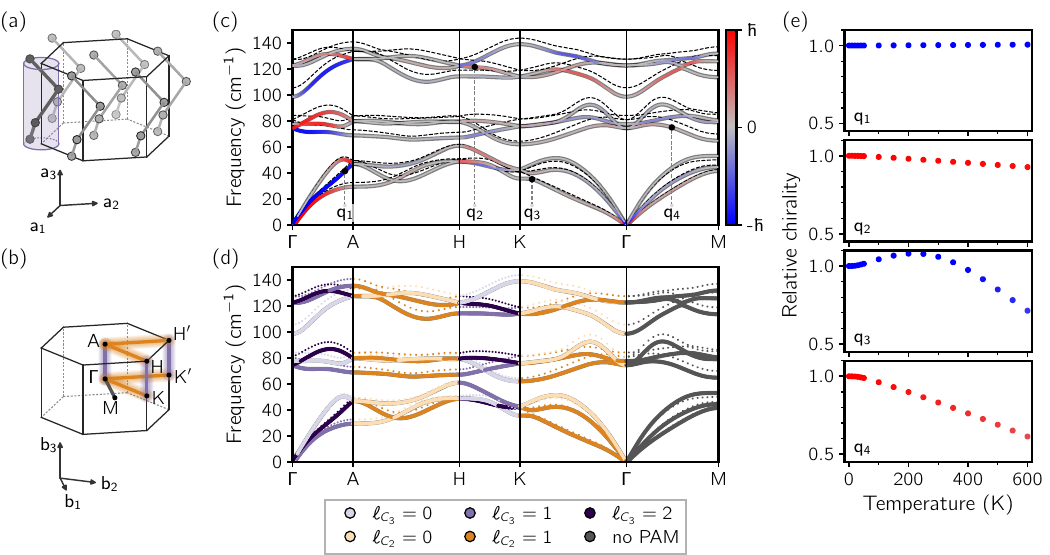}
    \caption{Anharmonic lattice dynamics in crystalline tellurium. (a) Structure and (b) Brillouin zone of right-handed crystalline tellurium. $C_3$ and $C_2$ rotationally invariant symmetry points (RISPs) are highlighted in purple and orange, respectively. (c) Harmonic dispersion relation (black dashed lines) and renormalized dispersion at 600\,K colored according to dynamic chirality, $\chi_{\bm{q}\lambda}$. (d) Spin pseudoangular momentum (PAM) of harmonic modes (dotted lines) and renormalized modes at 600\,K. (e) Relative temperature-dependent dynamic chirality of various modes at different points in the BZ (marked in panel (c)).}
    \label{fig:1}
\end{figure*}

In this work, we use an \textit{ab initio} many-body Green's function approach to investigate the lattice dynamics of crystalline tellurium (Te, Fig.~\ref{fig:1}a), which serves as a model helical crystal that is also of technological relevance~\cite{calavalle_gate-tuneable_2022, lin2016tellurium, tech2, tech3}, to understand the interplay between phonon chirality and anharmonicity. Namely, we include cubic and quartic anharmonicity to uncover 1) whether chiral vibrational modes persist at high temperatures, and 2) whether they live long enough to affect microscopic and macroscopic properties. We find that the discrete rotational symmetries of crystalline Te impart a symmetry label that restricts phonon-phonon interactions, such that the chirality of individual vibrational modes is robust to high temperature. Symmetry does not, however, protect lifetimes of chiral vibrational modes, leading to a significant reduction of lifetimes with increasing temperature, which has consequences for bulk angular momentum generation at elevated temperatures~\cite{zhang_measurement_2025}.  These results pave the way for furthering chiral materials as a platform for angular-momentum-based phononics.

The vibrational dynamics of a system are described by the Born-Oppenheimer potential energy surface (PES), which can be expanded with respect to atomic displacements from the equilibrium structure:
\begin{equation}
    U = U_0 + U_2 + U_3 + U_4 + \dots\,,
    \label{eqn:U_expansion}
\end{equation}
where the $N^{\text{th}}$ order contribution to the PES is given by
\begin{equation}
    U_N = \frac{1}{N!} \sum_{\{\bm{m},\alpha,i\}} \Phi^{(N)}_{\bm{m}_1\alpha_1i_1,\dots,\bm{m}_N\alpha_Ni_N} u_{\bm{m}_1\alpha_1i_1}\dots u_{\bm{m}_N\alpha_Ni_N}\,.
    \label{eqn:U_n}
\end{equation}
Here, $u_{\bm{m}\alpha i}$ denotes the displacement of atom $\alpha$ in unit cell $\bm{m}$ along the Cartesian direction $i$, and $\Phi^{(N)}$ represents the corresponding $N^{\text{th}}$-order interatomic force constant.

We begin by characterizing the lattice dynamics of crystalline Te within a harmonic description, which is determined by the dynamical matrix:
\begin{equation}
    \bm{\mathcal{D}}_{\alpha i,\beta j}(\bm{q}) = \frac{1}{\sqrt{m_\alpha m_\beta}} \sum_{\bm{m}}e^{i\bm{q}\cdot (\bm{R}_{\bm{m}\alpha} - \bm{R}_{\bm{0}\beta})}\Phi^{(2)}_{{\bm{m}\alpha i},\bm{0}\beta j}\,,
    \label{eqn:dynamical_matrix}
\end{equation}
where $\bm{q}$ is a wavevector in the first Brillouin zone (BZ), and $\bm{R}_{\bm{m}{\alpha}}$ is the equilibrium position of atom $\alpha$ in cell $\bm{m}$. We obtain the harmonic phonon modes, %$\bm{\epsilon}_\lambda(\bm{q})$ 
$\bm{\epsilon}_{\bm{q}\lambda}$, and frequencies, %$\omega_\lambda(\bm{q})$
$\omega_{\bm{q}\lambda}$, as the eigenvectors and the square root of the eigenvalues, respectively. The harmonic dispersion relation for Te is illustrated in the black dashed lines of Fig.~\ref{fig:1}c.

In crystals lacking inversion symmetry, such as Te, vibrational modes can exhibit finite microscopic angular momentum, or circular polarization (CP)~\cite{zhang_angular_2014}. Consequently, their atomic displacement patterns trace out circular or elliptical trajectories rather than the linear trajectories of conventional phonon modes. The CP of mode $\lambda$ at $\bm{q}$ is given by $\bm{s}_{\bm{q}\lambda}=\hbar\bm{\epsilon}^\dagger_{\bm{q}\lambda} \bm{\mathcal{S}}\bm{\epsilon}_{\bm{q}\lambda}$, where $\bm{\mathcal{S}}$ are the spin-1 matrices on a Cartesian basis, and takes on values between $-\hbar$ (left-handed) and $\hbar$ (right-handed). %$\bm{s}_\lambda(\bm{q})=\hbar\bm{\epsilon}^\dagger_\lambda(\bm{q}) \bm{\mathcal{S}}\bm{\epsilon}_\lambda(\bm{q})$
As transport phenomena are primarily sensitive to the CP component parallel to the direction of the propagation, we characterize the dynamic chirality~\cite{huang_2026} of each mode, %$\bm{s}_\lambda(\bm{q})\cdot\hat{\bm{q}}$
$\chi_{\bm{q}\lambda}=\bm{s}_{\bm{q}\lambda}\cdot\hat{\bm{q}}$, which provides a scalar measure of the handedness of a propagating vibrational mode.

In addition to generating physical angular momentum, these vibrational modes can be characterized according to their pseudoangular momentum (PAM)~\cite{zhang_chiral_2015, zhang_chiral_2022}. Analogous to the crystal momentum $\bm{q}$ that results from the discrete translational symmetry of periodic crystals, PAM is a symmetry label resulting from the discrete rotational symmetry of crystals like Te. Phonon PAM dictates scattering selection rules with quasiparticles, such as circularly polarized photons~\cite{ishito_truly_2023} and electrons~\cite{zhang_chiral_2015, zhang_chiral_2022}, as well as between phonons, as discussed further below. This work focuses on right-handed Te, which belongs to the $D_3$ point group and $P3_121$ space group, possessing one 3-fold screw symmetry axis along the helical chain ($C_3$) and three perpendicular 2-fold rotational axes ($C_2$). At rotationally invariant symmetry points (RISPs) of the BZ with $C_n\bm{q} = \bm{q}+\bm{G}$ (illustrated in Fig.~\ref{fig:1}b), the phonon Bloch states, %$\bm{u}_{\lambda,\bm{q}}(\bm{r})=\bm{\epsilon}_\lambda(\bm{q})e^{i\bm{q}\cdot\bm{R}_{\bm{m}\alpha}}$
$\bm{u}_{\bm{q}\lambda}(\bm{r})=\bm{\epsilon}_{\bm{q}\lambda}e^{i\bm{q}\cdot\bm{R}_{\bm{m}\alpha}}$, diagonalize the discrete rotation operator $\hat{C}_n$ with eigenvalues %$\exp[i\frac{2\pi}{3}l_\lambda(\bm{q})]$
$\exp[i\frac{2\pi}{n}l_{\bm{q}\lambda}]$. Here, %$l_\lambda(\bm{q})$
$l_{\bm{q}\lambda}$ is the PAM, which decomposes into spin and orbital contributions, %$l_\lambda(\bm{q})=l^{\text{s}}_\lambda(\bm{q})+l^{\text{o}}_\lambda(\bm{q})$
$l_{\bm{q}\lambda}=l^{\text{s}}_{\bm{q}\lambda}+l^{\text{o}}_{\bm{q}\lambda}$. The spin-PAM (Fig.~\ref{fig:1}d) takes on integer values between 0 and $n-1$ and is determined by %$\bm{\epsilon}_\lambda(\bm{q})$
$\bm{\epsilon}_{\bm{q}\lambda}$ while the orbital-PAM takes on continuous values and is determined by the Bloch phase, $e^{i\bm{q}\cdot\bm{R}_{\bm{m}\alpha}}$. 

While the harmonic approximation provides an accurate description of lattice dynamics at low temperatures, it fails to capture anharmonic effects, such as temperature-dependent frequencies, mode mixing, and finite lifetimes, that result from phonon-phonon interactions. These effects play a significant role in the lattice dynamics of Te~\cite{PhysRevB.4.356, PhysRevB.13.1383, lin2016tellurium, peng_anisotropic_2015, wu-prb2026}, which forms covalently bonded helical chains along the $c$-axis that are coupled only weakly through van der Waals interactions in the transverse directions~\cite{ghosh2008origin, yi2018nature}. The resulting contrast between strong intrachain covalent bonding and weak interchain coupling makes the lattice highly anisotropic and mechanically soft, producing pronounced deviations from harmonic behavior. In addition, the stereochemically active $5p$ lone pairs are highly polarizable, enhancing higher-order interatomic force constants and strengthening phonon-phonon interactions in a manner akin to SnSe~\cite{li2015orbitally}, certain perovskites~\cite{fabini2016dynamic, laurita2022chemistry}, and chalcogenide compounds~\cite{PhysRevLett.107.235901}.

To account for anharmonicity in the lattice dynamics, we calculate the anharmonic phonon Green's function (GF), $\bm{D}(\bm{q},\omega)$, using a first-principles approach. The phonon GF satisfies a Dyson equation,
\begin{equation}
    %\{\bm{D}(\bm{q},\omega)\}^{-1} = \{\bm{D}^0(\bm{q},\omega)\}^{-1} - \bm{\pi}(\bm{q},\omega)\,,
    \{\bm{D}_{\bm{q}}(\omega)\}^{-1} = \{\bm{D}^0_{\bm{q}}(\omega)\}^{-1} - 2\bm{\Omega}_{\bm{q}}\bm{\pi}_{\bm{q}}(\omega)\,,
    \label{eqn:Dyson}
\end{equation}
where %$\bm{D}^0(\bm{q},\omega)$
$\bm{D}^0_{\bm{q}}(\omega)=[\omega^2\bm{1}-\bm{\Omega}_{\bm{q}}^2]^{-1}$ and $\bm{\Omega}^2_{\bm{q}}$ are the harmonic GF and the dynamical matrix, respectively, and %$\bm{\pi}(\bm{q},\omega)$
$\bm{\pi}_{\bm{q}}(\omega)$ is the phonon self-energy that describes anharmonicity. The most important contributions to the self-energy from third- and fourth-order anharmonicity are the bubble and loop diagrams, respectively.

The loop diagram accounts for temperature-dependent frequency shifts and mode-mixing due to the thermal occupation of phonons. Self-consistent phonon (SCP) theory~\cite{Hooton1958,Cowley1963,Tadano_jpsj_2018} incorporates this contribution nonperturbatively by iteratively renormalizing the anharmonic frequencies and modes, thereby summing the loop diagrams to infinite order. The corresponding Dyson equation for this renormalized Green's function, %$\bm{D}^S(\bm{q},\omega)$
$\bm{D}^S_{\bm{q}}(\omega)$, is
\begin{equation}
    %\{\bm{D}^S(\bm{q},\omega)\}^{-1} = \{\bm{D}^0(\bm{q},\omega)\}^{-1} - \bm{\pi}^L[\bm{D}^S(\bm{q},\omega)]\,,
    \{\bm{D}^S_{\bm{q}}(\omega)\}^{-1} = \{\bm{D}^0_{\bm{q}}(\omega)\}^{-1} - 2\bm{\Omega}_{\bm{q}}\bm{\pi}^L_{\bm{q}}[\bm{D}^S_{\bm{q}}(\omega)]\,,
    \label{eqn:Dyson_SCP}
\end{equation}
where the loop self-energy,
\begin{align}
    %\bm{\pi}^L_{\lambda,\lambda'}(\bm{q}) = -\frac{\hbar}{2N_q}\sum_{\bm{q}_1,\lambda_1}&\frac{ V^{(4)}_{\lambda,\lambda',\lambda_1,\lambda_1}(\bm{q}, -\bm{q}, \bm{q}_1, -\bm{q}_1)}{4\omega_{\lambda_1}(\bm{q}_1)\sqrt{\omega_\lambda(\bm{q})\omega_{\lambda'}(\bm{q})}} \nonumber \\
    %&\times [1+2n(\omega_{\lambda_1}(\bm{q}_1)]\,,
    \bm{\pi}^L_{\bm{q}\lambda,\lambda'} = -\frac{\hbar}{2N_q}\sum_{\bm{q}_1\lambda_1}&\frac{ V^{(4)}_{\bm{q}\lambda,-\bm{q}\lambda',\bm{q}_1\lambda_1,-\bm{q}_1\lambda_1}}{4\sqrt{\omega_{\bm{q}\lambda}\omega_{\bm{q}\lambda'}}\omega_{\bm{q}_1\lambda_1}}[1+2n(\omega_{\bm{q}_1\lambda_1})]\,,
    \label{eqn:loop}
\end{align}
is evaluated self-consistently. Here, $N_q$ is the number of $\bm{q}$ points in the first BZ, $V^{(4)}$ is the fourth order force constant Fourier transformed to the harmonic phonon basis, and $n(\omega)$ is the Bose-Einstein distribution. SCP yields temperature-dependent renormalized vibrational modes, where %$\Omega_{\mu}(\bm{q})$ 
$\omega^S_{\bm{q}\mu}$ denotes the frequency of the renormalized mode $\mu$ at $\bm{q}$.

The bubble diagram represents three-mode scattering, and its inclusion leads to the following approximation to the Dyson equation:
\begin{equation}
    %\{\bm{D}(\bm{q},\omega)\}^{-1} \approx \{\bm{D}^S(\bm{q},\omega)\}^{-1} - \bm{\pi}^B[\bm{D}^S(\bm{q},\omega)]\,.
    \{\bm{D}_{\bm{q}}(\omega)\}^{-1} \approx \{\bm{D}^S_{\bm{q}}(\omega)\}^{-1} - 2\bm{\Omega}^S_{\bm{q}}\bm{\pi}^B_{\bm{q}}[\bm{D}^S_{\bm{q}}(\omega)]\,,
    \label{eqn:Dyson_bubble}
\end{equation}
where $\bm{\Omega}^S_{\bm{q}}$ now includes the SCP contribution to the dynamical matrix, and the bubble self-energy is evaluated in the temperature-dependent renormalized basis as
\begin{align}
    %\bm{\pi}^B_{\mu,\mu'}(\bm{q}, \omega) = \frac{\hbar}{2N}\sum^\prime_{\substack{\bm{q}_1,\bm{q}_2\\\mu_1,\mu_2}}&\frac{ V^{(3)}_{\mu,\mu_1,\mu_2}(-\bm{q},\bm{q}_1,\bm{q}_2)V^{(3)}_{\mu',\mu_1,\mu_2}(\bm{q},-\bm{q}_1,-\bm{q}_2)}{8\sqrt{\Omega_{\mu}(\bm{q})\Omega_{\mu'}(\bm{q})}\Omega_{\mu_1}(\bm{q}_1)\Omega_{\mu_2}(\bm{q}_2)} \nonumber \\
    %&\times \mathcal{F}_{\mu_1,\mu_2}(\omega,\bm{q}_1,\bm{q}_2)\,,
    \bm{\pi}^B_{\bm{q}\mu,\mu'}(\omega) = \frac{\hbar}{2N_q}\sum^\prime_{\substack{\bm{q}_1\mu_1\\\bm{q}_2\mu_2}}&\frac{ V^{(3)}_{-\bm{q}\mu,\bm{q}_1\mu_1,\bm{q}_2\mu_2}V^{(3)}_{\bm{q}\mu',-\bm{q}_1\mu_1,-\bm{q}_2\mu_2}}{8\sqrt{\omega^S_{\bm{q}\mu}\omega^S_{\bm{q}\mu'}}\omega^S_{\bm{q}_1\mu_1}\omega^S_{\bm{q}_2\mu_2}} \nonumber \\
    &\times \mathcal{F}_{\bm{q}_1\mu_1,\bm{q}_2\mu_2}(\omega)\,.
    \label{eqn:bubble}
\end{align}
Here, the primed summation indicates only $(\bm{q}_1,\bm{q}_2)$ pairs satisfying momentum conservation are included, and %$\mathcal{F}_{\mu_1,\mu_2}(\omega,\bm{q}_1,\bm{q}_2)$
$\mathcal{F}_{\bm{q}_1\mu_1,\bm{q}_2\mu_2}(\omega)$ is a frequency-dependent function. Within the relaxation time approximation, %an additional frequency shift is given by %$\Delta_\mu(\bm{q}) = -\Re[\bm{\pi}^B_{\mu,\mu}(\bm{q},\Omega_\mu(\bm{q}))]/\hbar$
%$\Delta_{\bm{q}\mu} = -\Re[\bm{\pi}^B_{\bm{q}\mu}(\omega^S_{\bm{q}\mu})]$
the frequency shifts to $\big[(\omega_{\bm{q}\mu}^S)^2 + 2\omega_{\bm{q}\mu}^S\Re\bm{\pi}^B_{\bm{q}\mu}(\omega^S_{\bm{q}\mu})\big]^{1/2}$, and the lifetime is given by %$\tau^{-1}_\mu(\bm{q}) = 2\Im[\bm{\pi}^B_{\mu,\mu}(\bm{q},\Omega_\mu(\bm{q}))]/\hbar$
$\tau^{-1}_{\bm{q}\mu} = 2\Im[\bm{\pi}^B_{\bm{q}\mu}(\omega^S_{\bm{q}\mu})]$. Note that we only consider diagonal elements of the bubble self-energy, as the quasiparticles defined by the SCP approach are appropriate~\cite{Tadano_prl_2022}. Combining SCP theory with the leading-order bubble correction, which we refer to as SCP+B, thus provides an efficient first-principles framework for describing anharmonic lattice dynamics that simultaneously captures nonperturbative frequency renormalization, mode-mixing, and finite phonon lifetimes. All computational details regarding interatomic force constants, SCP, and SCP+B calculations are provided in the SM.

We calculate the anharmonic phonon GF and illustrate the SCP+B renormalized dispersion relation at 600\,K in Fig.~\ref{fig:1}c, which shows significant frequency renormalization of about 10--15\%, comparable to the renormalization found in PbTe and related materials~\cite{PhysRevB.91.214310, PhysRevB.97.014306}. We find that the dynamic chirality of these renormalized modes is robust at 600\,K (colorbar in Fig.~\ref{fig:1}c) throughout the BZ, especially along the RISPs on the $\Gamma\text{-}A$ and $H\text{-}K$ paths. Similarly, Fig.~\ref{fig:1}d demonstrates that spin-PAM is totally preserved at high temperature and that renormalized modes retain the same integer values as the corresponding harmonic modes. These results indicate that anharmonic lattice dynamics retain their chirality to high temperatures and are robust to the effects of mode-mixing.

The preservation of chirality in the anharmonic lattice dynamics can be understood by an analysis of the scattering selection rules, which can be derived using symmetry constraints imposed on the anharmonic interaction vertices. As we show in the SM, the $N$-th order anharmonic interaction vertex, $V_{\lambda_1,\ldots,\lambda_N}^{(N)}
(\bm q_1,\ldots,\bm q_N)$, is non-zero only when $\bm q_1 + \dots + \bm q_N =\bm G$ and $\left[l_{\bm{q}_1\lambda_1}^{\text{s}} + \dots + l_{\bm{q}_N\lambda_N}^{\text{s}} +l_G\right] \bmod n=0$. The first of these conditions is the well-known conservation of crystal momentum, which holds for interactions between any $N$ phonons in the BZ. The second condition enforces the conservation of spin-PAM, which only applies to interactions between phonons at RISPs for which spin-PAM is well defined. We note that the selection rules shown here are consistent with those determined from group theoretic and symmetry-based approaches in other systems~\cite{cammarata2019, liao2021, lindsay2022, mingo2009, giantthermalconductivity2026}.

The consequences of this additional selection rule become evident in both perturbative and nonperturbative treatments of anharmonicity. Within SCP theory, the loop contribution to the self-energy depends on the quartic vertex (Eq.~\ref{eqn:loop}), which requires
\begin{equation}
    [l^{\text{s}}_{\bm{q}\lambda} -l^{\text{s}}_{\bm{q}\lambda'}]\bmod n =0\,.
    \label{eqn:loop_selectionRule}
\end{equation}
This result indicates that although every thermally occupied mode contributes to frequency renormalization, SCP only mixes modes of the same spin-PAM, and so spin-PAM is a robust symmetry label beyond the harmonic approximation. 

Furthermore, Eq.~\ref{eqn:loop_selectionRule} limits mode-mixing along RISPs (Fig.~S2 in the SM) so that those modes retain their CP and dynamic chirality (Fig.~\ref{fig:1}c). We investigate the temperature-dependence of dynamic chirality of various acoustic and optical modes in the BZ. As shown in Fig.~\ref{fig:1}e, the dynamic chirality at ${C}_3$ RISPs is robust to high temperature, with the modes at $\bm{q}_1$ and $\bm{q}_2$ maintaining 100\% and 93\% of their harmonic values at 600\,K, respectively. The mode at $\bm q_3$ is along a ${C}_2$ RISP and mixes with another mode of the same spin-PAM  to form an anharmonic avoided crossing. As a result, the CP of this mode ``constructively" interferes with its mixing counterpart to realize a dynamic chirality that is larger than its harmonic value at intermediate temperatures around 200\,K and then ``destructively" interferes to a dynamic chirality that is 73\% of its harmonic value at 600\,K. Away from any RISP, the mode at $\bm{q}_4$ mixes readily with other modes, and the dynamic chirality decays more significantly with temperature, dropping to 62\% of its harmonic value at 600\,K. While the conservation of spin-PAM is only well-defined at the RISPs, vibrational modes at these symmetry-protected regions are those with the strongest CP. Further, spin-PAM is approximately conserved at points near the RISPs, and the robustness of dynamic chirality extends to these modes as well (Fig.~S5 of the SM). Together, these results highlight the broader influence of rotational symmetry on anharmonic chiral lattice dynamics.

\begin{figure}[t]
    \centering
    \includegraphics[width=\columnwidth]{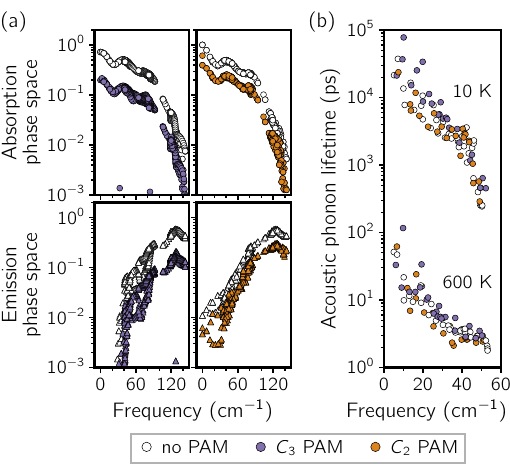}
    \caption{The effects of spin-PAM symmetry selection rules on three-mode scattering. (a) The three-mode absorption (top) and emission (bottom) scattering phase space of triplets with all three modes along ${C}_3$ RISPs (purple) and ${C}_2$ RISPs (orange). Accounting for the conservation of spin-PAM results in a 3-fold and 2-fold reduction, respectively, in the scattering phase space. (b) Temperature-dependent lifetimes of acoustic modes along the $\Gamma$-$A$ path (purple, $C_3$ invariant), $\Gamma$-$K$ path (orange, $C_2$ invariant), and $\Gamma$-$M$ path (white, no symmetry invariance).}
    \label{fig:2}
\end{figure}

The conservation of spin-PAM also affects three-mode scattering, which is described in our SCP+B approach by the cubic vertices of the bubble contribution to the self-energy:
\begin{align}
\left[l_{\bm q\mu}^{\text{s}}-l_{\bm q_1\mu_1}^{\text{s}}-l_{\bm q_2\mu_2}^{\text{s}}+ l_G\right]\bmod n&=0, \nonumber\\\left[l_{\bm q\mu}^{\text{s}}-l_{\bm q\mu'}^{\text{s}}\right]\bmod n&=0
    \label{eqn:bubble_selectionRule}\,.
\end{align}
As illustrated in Fig.~\ref{fig:2}a, accounting for the conservation of spin-PAM (in addition to the usual conservation of energy and crystal momentum) reduces the three-mode scattering phase space significantly. The interaction of modes along $C_3$ and $C_2$ RISPs show a reduction to 1/3 and 1/2, respectively, in both the absorption and emission scattering phase spaces. Because $n$-fold rotational symmetry results in $n$ distinct PAM values, the three-mode on-axis scattering phase space is reduced by a factor of $n$. When using symmetry-enforced force constants, as done in this work, these selection rules do not need to be explicitly enforced, but emerge intrinsically from the Fourier-transformed force constants (Fig.~S3 of the SM)~\cite{lindsay2022}. Despite this reduction in the scattering phase space, the bubble self-energy involves a summation over all three-mode scattering processes (Eq.~\ref{eqn:bubble}) such that the effects of symmetry on the mode lifetimes are less pronounced. Fig.~\ref{fig:2}b shows the lifetimes of acoustic modes along the $\Gamma$-$A$ path ($C_3$ invariant), $\Gamma$-$K$ path ($C_2$ invariant), and $\Gamma$-$M$ path (no symmetry invariance, PAM undefined). At 10\,K, the lifetimes of the $C_3$ invariant acoustic mode are slightly extended, but attributing these longer lifetimes to a symmetry-based reduction in scattering is challenging. Any potential effects are washed out at 600\,K, when scattering rates are drastically increased and the lifetimes of all acoustic modes are comparable with one another. 

Thus, despite retaining their microscopic chiral character at elevated temperatures, these anharmonic chiral modes are relatively short lived; still, they remain underdamped, undergoing approximately 4–6 oscillation cycles within their lifetimes at 600\,K. This contrasting behavior can be understood from the structure of the anharmonic phonon self-energy, which is block diagonal in the spin-PAM basis at RISPs, as symmetry permits nonzero non-diagonal elements only between modes belonging to the same spin-PAM (Eqs.~\ref{eqn:loop_selectionRule} and \ref{eqn:bubble_selectionRule}). Even within these symmetry-allowed blocks, non-diagonal elements are generally suppressed by the increasing frequency separation between branches of the same spin-PAM. As mode-mixing is dictated by non-diagonal elements of the self-energy, spin-PAM conservation protects the CP and dynamic chirality of anharmonic modes by restricting mode mixing. Frequency renormalization and lifetimes, however, are governed by diagonal elements of the self-energy, where the spin-PAM selection rule is trivially satisfied and thus symmetry restriction has a negligible effect. We emphasize that the form of the self-energy, and its consequences on anharmonic mode-mixing, frequency renormalization, and lifetimes, is general and applies even beyond the SCP+B framework (additional discussion in SM).

This dichotomy has a direct macroscopic consequence on the total phonon angular momentum of the crystal, $\bm{J}_{\text{ph}}$, which is related to the Bose occupation and circular polarization of each mode~\cite{zhang_angular_2014}. While $\bm{J}_{\text{ph}}$ is zero in equilibrium, it can be generated through time-reversal symmetry breaking and thereby induce magnetization of the material or result in rigid-body rotation of the crystal~\cite{hamada_phonon_2018,qin_chiral_2025,zhang_measurement_2025}. The total phonon angular momentum under the application of a temperature gradient is given by $J_{\text{ph},i}=\alpha_{ij}\nabla_jT$~\cite{hamada_phonon_2018}, where the response tensor, $\bm{\alpha}$, is a material property that relates the applied temperature gradient to the generated angular momentum current. Understanding $\bm{\alpha}$ as well as its temperature-dependent behavior and tunability is a topic of current interest~\cite{hamada_phonon_2018,qin_chiral_2025,pols_chiral_2025, zhang_measurement_2025, pols2025}.

While anharmonic phonon-phonon scattering fundamentally underpins a finite-valued phonon angular momentum response tensor, past calculations of $\bm{\alpha}$ have relied on a harmonic approximation and the constant relaxation time approximation (C-RTA), in which all modes are governed by a single phenomenological, temperature-independent lifetime, $\tau$. As a result, the temperature-dependence of $\bm{\alpha}$ only enters through the thermal occupation of modes and is unable to qualitatively explain high temperature experimental results~\cite{zhang_measurement_2025}. Here, we present the determination of $\bm{\alpha}$ that incorporates mode-resolved, temperature-dependent lifetimes, which are microscopically determined by our SCP+B approach, as well as temperature-dependent frequencies, group velocities, and circular polarizations:
\begin{equation}
    \alpha_{ij}(T) = \frac{-1}{N_qV}\sum_{\bm{q}\mu} s_{\bm{q}\mu,i}(T) v_{\bm{q}\mu,j}(T)\tau_{\bm{q}\mu}(T)\frac{\partial n\big(\omega_{\bm{q}\mu}(T)\big)}{\partial T}\,.
    \label{eqn:alpha_T}
\end{equation}

\begin{figure}[t]
\centering
\includegraphics[width=\columnwidth]{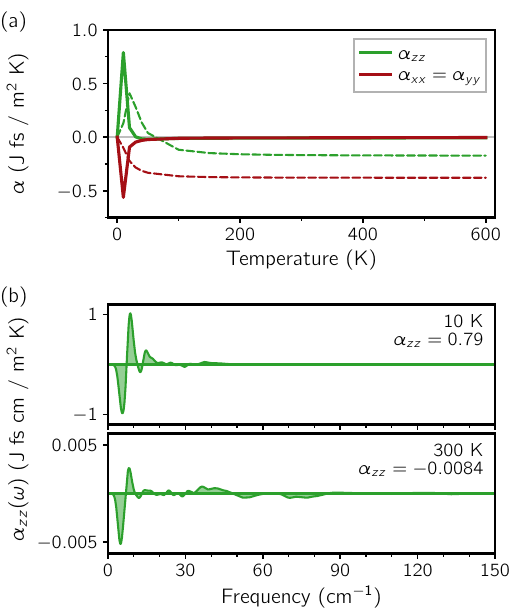}
\caption{Temperature-dependent angular momentum generation. (a) Temperature-dependent phonon angular momentum response tensor calculated using the harmonic constant relaxation time approximation (dashed lines) and using microscopically temperature-dependent quantities, according to Eq.~\ref{eqn:alpha_T}. (b) Spectral decomposition of $\alpha_{zz}$ at 10\,K (top) and 300\,K (bottom).}
\label{fig:3}
\end{figure}

Fig.~\ref{fig:3}a illustrates the temperature-dependence of the phonon angular momentum response tensor computed within the harmonic C-RTA approximation with a single lifetime of $\tau=1$\,ns (approximated from our calculated phonon lifetimes at 10\,K in Fig.~\ref{fig:2}b). At all temperatures, the only non-zero values of the response tensor are $\alpha_{xx}=\alpha_{yy}$ and $\alpha_{zz}$, as dictated by the $D_3$ point group of Te~\cite{hamada_phonon_2018}. The harmonic C-RTA approximation indicates that $\alpha_{zz}$ changes sign with increasing temperature, as dispersive optical modes with negative dynamic chirality become more populated, but that it does not decrease significantly in magnitude. These calculations are consistent with theoretical predictions within the C-RTA~\cite{hamada_phonon_2018}. The values of $\bm{\alpha}$ change significantly once anharmonicity is microscopically introduced and the response tensor is computed using temperature-dependent quantities as given by Eq.~\ref{eqn:alpha_T}, causing a reduction in $\alpha_{zz}$ from a value of $0.790$\,J$\cdot$fs$\cdot$m$^{-2}\cdot$K$^{-1}$ at 10\,K to $-0.008$\,J$\cdot$fs$\cdot$m$^{-2}\cdot$K$^{-1}$ at 300\,K to $-0.005$\,J$\cdot$fs$\cdot$m$^{-2}\cdot$K$^{-1}$ at 600\,K. As shown in Fig.~S4 in the SM, the drop in mode lifetimes with increasing temperature is the dominating factor in reduction of the response tensor $\bm \alpha$, whereas, the temperature-dependence of the mode frequencies, CPs, and group velocities have a minimal effect.

We further examine the angular momentum generation through a spectral decomposition, where $\bm{\alpha}(\omega,T) = \sum_{\bm{q}\mu}\bm{\alpha}_{\bm{q}\mu}(T)\delta\big(\omega-\omega_{\bm{q}\mu}(T)\big)$. Fig.~\ref{fig:3}b demonstrates that the largest contributions to the generated angular momentum come from the transverse acoustic (TA) modes due to their strong CP and large group velocities. The first, lower-energy TA mode has a negative contribution while the second, higher-energy TA mode has a positive contribution. As the temperature increases, higher-frequency modes become populated, shifting the overall value of $\alpha_{zz}$ from positive to negative, especially with the population of the dispersive optical bands between 100--120\,cm$^{-1}$.

Our results are consistent with the experimental measurements of Zhang \textit{et al.}~\cite{zhang_measurement_2025}, who observed a macroscopic torque, which is proportional to the generated angular momentum, in response to an applied thermal gradient along the chiral axis in single-crystal Te. The torque was experimentally attributed to chiral phonon activation at 10\,K but was dominated by thermal expansion at 300\,K. This high-temperature observation contrasts with predictions within the C-RTA (Fig.~\ref{fig:3}a), which suggest a phonon-induced torque of comparable magnitude but opposite sign at 300\,K. Our anharmonic calculations reveal that the generated angular momentum at elevated temperatures is $~100\times$ smaller and of opposite sign due to the short lifetimes of chiral modes at those temperatures (Fig.~\ref{fig:3}b), consistent with a signal dominated by extrinsic effects.

In conclusion, we have used a first-principles many-body Green's function approach to investigate the interplay of chirality and anharmonicity in the lattice dynamics of crystalline Te. We find that discrete rotational symmetries in the crystal structure impose selection rules that restrict mode mixing along RISPs such that the dynamic chirality of vibrational modes remains robust at elevated temperatures. Importantly, the modes that are symmetry protected are also those that have the most chiral character and that dominate angular momentum generation. While symmetry selection rules also limit three-mode scattering, they have a minimal impact on mode lifetimes.
As a result, the macroscopic phonon angular momentum response decreases strongly with temperature, providing insight into recent experimental observations. Despite the finite lifetimes of chiral vibrational modes, they remain underdamped even at 600\,K, and can participate in electron-phonon and spin-phonon processes. The temperature-independent symmetry selection rules identified here will also constrain these interactions, potentially influencing chiral phonon-mediated orbital and spin transport. In addition to the generation of bulk angular momentum, a nonequilibrium population of phonons may also generate a net PAM that transfers to electrons, exerting a ``chiral phonon drag" on the electronic system that increases the spin-Seebeck effect, analogous to the standard phonon drag effect~\cite{maekawa2010, kim_chiral-phonon-activated_2023}. Finally, the present work highlights the importance of the RISPs in protecting chiral quasiparticle properties, suggesting that future efforts in BZ engineering to increase the relative volume of symmetry-protected regions of the BZ, such as through the design of low-dimensional, heterostructured, or hybrid materials, may be a successful route toward enhancing chiral properties.

{\it{Acknowledgement:}} D.J. acknowledges start-up funding support from the University of Chicago. This work was completed using resources from the University of Chicago's Research Computing Center. 

%\bibliography{refs}

\end{document}